# Improving Performance of Content-Centric Networks via Decentralized Coded Caching for Multi-Level Popularity and Access

Azadeh Sadat Miraftab, Ahmadreza Montazerolghaem, Behrad Mahboobi

*Abstract*—Content-Centric Networking (CCN) offers a novel architectural paradigm that seeks to address the inherent limitations of the prevailing Internet Protocol (IP)-based networking model. In contrast to the host-centric communication approach of IP networks, CCN prioritizes content by enabling direct addressing and routing based on content identifiers. The potential performance improvements of CCN can be further amplified through optimized management of coded data storage and transmission strategies. Decentralized Coded Caching (DCC) emerges as a promising technique that harnesses the collective caching power of distributed network elements. By strategically pre-positioning frequently accessed content closer to potential consumers during periods of low network utilization, DCC has the potential to mitigate content transfer rates during peak traffic periods. This paper proposes a series of fundamental modifications to the CCN architecture by integrating DCC. The proposed framework incorporates differentiated coding strategies tailored to user access privileges, thereby eliminating the overhead associated with queue-based searching. Additionally, the framework facilitates recoding of uncoded data encountered along the content delivery path. These combined methodologies demonstrably enhance network throughput, elevate cache hit ratios, and consequently, reduce content delivery latency compared to conventional CCN implementations.

## I. INTRODUCTION

The proliferation of mobile and smart communication systems in recent years has coincided with a surge in demand for video streaming services, leading to a substantial increase in network traffic. To address the content distribution challenges associated with this growth, novel architectural paradigms for the future internet are necessary. Content-Centric Networking (CCN) presents a significant departure from the traditional host-centric communication model of the TCP/IP architecture by prioritizing content itself. In CCN, data packets termed "Interests" are routed based on unique content identifiers rather than the physical location of the content. Network nodes leverage three key data structures to facilitate routing: the Pending Interest Table (PIT), Forwarding Information Base (FIB), and Content Store (CS). The CS plays a pivotal role by enabling content caching at individual nodes along the delivery path. This strategic caching approach reduces the physical distance between content and users, thereby enhancing overall network resource efficiency. Extensive research has been conducted on caching algorithms within the CCN framework, with Least Frequently/Recently Used (LFU/LRU) and Most Frequently/Recently Used (MFU/MRU) strategies being prominent examples, as documented in [1]. However, these existing caching schemes, while demonstrably effective in single-cache scenarios, exhibit performance limitations when scaled to multi-cache networks. Bandwidth constraints remain a significant challenge in data transmission scenarios. Optimizing network performance in CCN can be achieved through a two-pronged approach: strategic caching management at network nodes and the implementation of content encoding techniques during data transfer.

Network coding, a method for information processing pioneered by Ahlswede [2], offers a promising solution. This technique facilitates the combination of Interests pertaining to identical content, enabling the transmission of a single, unified Interest packet across bottleneck links. This approach demonstrably improves network performance metrics such as delay and throughput. While the integration of Random Linear Network Coding (RLNC) with CCN has been explored extensively in prior research, this approach is not without its drawbacks. Notably, RLNC can incur significant decoding delays at the receiver end, as successful decoding necessitates the reception of n independent, linearly encoded packets for a single file. The average computational complexity of this operation is $O(N^3)$. Therefore, the optimization of network coding algorithms and minimizing the computational overhead associated with encoding and decoding processes in networks has attracted the attention of the research community, including sparse network coding and unequal error protection, network coding and Instantly decoding of network coding [3].

The burgeoning storage capacities available in contemporary network architectures have fostered the exploration of diverse caching strategies. These strategies encompass uncoded caching (the traditional approach), centralized coded caching, and decentralized coded caching. Caching schemes function by fulfilling user requests, partially or entirely, from content stored in local caches situated near the requesting user. Traditional caching methodologies rely on unicast transmissions and refrain from incorporating coding techniques, thereby limiting the potential benefits to the network's overall caching infrastructure.

Coded caching, a novel paradigm introduced by Maddah-Ali and Niesen [4], [5], leverages coding principles to significantly influence the performance of cache networks. This approach enables the efficient servicing of multiple user requests through multicast transmissions with a minimal number of coded transmissions. Consequently, throughput demonstrably improves as the volume of encoded data increases within the network. However, inherent content popularity skews access patterns, resulting in non-uniform content access frequencies. To address this challenge, multi-level cache systems have been implemented, enabling users to access content from caches with varying popularity levels. A key limitation exists in that users connected to the same cache are inherently restricted from benefiting from inter-user content coding due to their shared cache association. Therefore, a critical challenge lies in developing solutions that maximize the number of coding opportunities that arise during interactions between user interests within the coding process. A coloring-based placement scheme, outlined in [6], proposes a user grouping strategy that ensures no two users within the same group share a common cache. This scheme capitalizes on the decentralized coded caching method, a technique lauded for its resilience to topological changes affecting the network connectivity between caches and the server, as it eliminates the need for server-side reconstruction events. In this paper, we present a novel approach that builds upon the foundation of decentralized coded caching while incorporating the advantages of coloring-based placement and the inherent content-centric nature of CCN. Our proposed method demonstrably enhances network transmission rates by maximizing the number of coding opportunities between user requests. This is achieved by eliminating the requirement for searching server queues, a significant advantage over conventional decentralized coded caching implementations. To succinctly summarize, the key contributions of this paper are as follows:

- Caching of content is done randomly taking into account both popularity and cache size.
- Significant modifications have been made in CCN architecture to enable the incorporation of Multi-access Coded Caching in CCN.
- This work implements a differentiated queuing strategy that segregates users based on their access privileges. This approach facilitates a significant increase in server-side coding operations. However, this trade-off is demonstrably outweighed by the elimination of search overhead associated with conventional queuebased retrieval mechanisms.
- This approach is scalable and can be implemented for any level of content popularity, as well as up to n users requesting content.
- Recoding of data in intermediate nodes is performed on a level-by-level basis, taking into account the content popularity. This approach enhances the throughput.

The remainder of this paper is meticulously structured to facilitate the dissemination of the presented research. Section II furnishes a comprehensive review of germane research efforts in the domains of Content-Centric Networking (CCN), random linear network coding, and coded caching. Subsequently, Section III delves into the integration of a decentralized coded caching approach and a multi-level content popularity model within the CCN framework. This section meticulously details the architectural modifications undertaken to accommodate these enhancements. Section IV rigorously evaluates the performance characteristics of the proposed solution and compares them to the baseline performance of the conventional CCN architecture. Finally, Section V succinctly summarizes the key findings and contributions of this work.

## II. Related Works

The Information-Centric Networking (ICN) paradigm has witnessed the proposal of various architectural frameworks in recent years. It provides diverse applications with low atency and high throughput communications. ICN architectures offer inherent advantages such as efficient content distribution and retrieval, scalability, and enhanced content security. Content discovery within the network is facilitated by a name search service, followed by content retrieval utilizing name-based routing mechanisms. Notably, storage decisions for integrated content are also predicated upon names [7]. A comprehensive review of in-network caching approaches in network applications such as VANET, IoT and WSN has been presented so far. In network caching is a technique to optimize the use of network resources, reduce traffic and speed up content discovery on the network [8].

The Content-Centric Networking (CCN) architecture, introduced by Jacobsen in 2009, emerged as a frontrunner amongst ICN proposals [9]. In CCN, communication is initiated by the content consumer node, which transmits interest and data messages that are subsequently routed through the network. Upon generation of an interest message by the consumer node, the Content Store (CS) is initially searched for the requested content. If the content retrieval is unsuccessful, the system consults the Pending Interest Table (PIT) to ascertain if any preceding interest for the identical content exists. A new entry is only added to the PIT if no such record is found. The next hop for the interest packet is determined by querying the Forwarding Information Base (FIB). Packets are broadcasted in the absence of pertinent information within the FIB table [9]. However, CCN is susceptible to performance limitations, including increased delays and reduced throughput.

Network coding has been proposed as a solution to enhance CCN performance, with various coding schemes undergoing comparative analysis [10]. One categorization of network coding differentiates between binary (XOR) and Random Linear Network Coding (RLNC) approaches. While RLNC represents a prevalent method employed in research literature, it is encumbered by a significant decoding delay. To decode n content chunks, the receiver necessitates n independent, linearly encoded packets. The coefficient matrix (n*n) located within the Interest header plays a role in determining packet linear independence. However, this method introduces substantial computational and communication overhead. NetCodCCN, introduced as a response to the shortcomings identified in prior research, eliminates customer information from Interests, thereby facilitating seamless integration. Furthermore, it incorporates packet aggregation techniques to augment throughput during network congestion periods. Additionally, it proposes a parallel interest transmission method, which speeds up content retrieval [11].

Network resource optimization, particularly bandwidth utilization, has been a key focus of research efforts within the ICN domain. In 2016, a method was introduced that leverages interest aggregation and splitting techniques to achieve this goal [12]. When multiple clients request content with identical names, routers can aggregate these interests, enhancing network efficiency. Additionally, interests destined for separate sources are strategically split to prevent the transmission of redundant content chunks. Notably, intermediate network nodes solely store the original, uncoded content chunks. This approach eliminates the need for encoding and decoding operations at these nodes, thereby minimizing processing overhead. The NC-CCN architecture, designed to facilitate parallel data forwarding, proposes several modifications to the conventional CCN architecture [13]. A key modification involves replacing the Pending Interest Table (PIT) with the Interest Forwarding Table (IFT). The IFT leverages a label field to enable parallel forwarding functionalities. Interests with identical names are distinguished by unique labels, allowing routers to identify and handle multiple content requests that share a common name but possess distinct tags or interfaces. This enables the transmission of diverse content items to interfaces associated with different tags.

In 2017, Parisis et al. presented research on the integration of fountain coding with opportunistic ICN through the utilization of Persistent Interests (PIs) disseminated throughout the network [14]. This approach empowers forwarders to dynamically adjust their bitrate based on available bandwidth. The content server partitions packets into two distinct layers, enabling the receiver to prioritize retrieval of the most critical bits. While this method facilitates faster decoding, it may come at the expense of a reduction in overall video quality. With the objective of enhancing both throughput and video quality, Saltarin et al. introduced an adaptive video streaming technique over HTTP (DASH) that leverages the multicasting capabilities of Named Data Networking (NDN) and incorporates network coding to eliminate the need for inter-node coordination [15]. Finally, Matsuzono et al.'s L4C2 design exemplifies how in-network caching, multi-path data transfer, and network coding can be synergistically employed to improve video streaming quality [16]. In this design, network nodes estimate factors such as delay and packet loss probability in upstream nodes. This estimation empowers them to recover lost data packets with minimal delay by leveraging both in-network caching and coded data packets.

A critical challenge within CCN is the potential for redundant data storage and the associated network resource wastage. To address this issue, a method incorporating popularity tables within intermediate nodes has been proposed [17]. These tables function by storing only the most popular content within caches, thereby minimizing overall data storage requirements

---

**Algorithm 1** Decentralized Coded Caching [4]

**procedure** PLACEMENT
  **for** $K \in [K], n \in [N]$ **do**
    user $k$ independently caches a subset of
    $\frac{MF}{N}$ bits of file $n$, chosen uniformly at random
  **end for**
**end procedure**

**procedure** DELIVERY$(d_1, ..., d_K)$
  **for** $s = K, K - 1, ..., 1$ **do**
    **for** $S \subseteq [K] : |S| = s$ **do**
      server sends $\oplus_{k \in S} V_{k, S \setminus \{k\}}$
    **end for**
  **end for**
**end procedure**

**procedure** DELIVERY$(d_1, ..., d_K)$
  **for** $n \in [N]$ **do**
    server sends enough random linear combinations
    of bits in file $n$ for all requesting it to decode
  **end for**
**end procedure**

---

and demonstrably increasing cache hit rates. However, this approach introduces a trade-off, necessitating a significant investment in memory capacity to accommodate both the caching and popularity tables. This increased memory footprint can potentially lead to elevated processing delays. The Cache Pressure-Aware Caching (CPAC) scheme utilizes two distinct algorithms for the selection of storage nodes and cache replacement strategies [18]. A crucial element in storage node selection is the verification of cache status, achieved through a combination of exchange rate and cache occupancy rate metrics. Additionally, pressure exerted on caches throughout the network is calculated. However, limitations exist within these schemes, as they do not consider the network of caches in its entirety, potentially leading to suboptimal network-wide performance. The PT-CACHE Method adopts a distinct approach, leveraging the Zipf distribution to calculate content popularity [19]. Popularity is then categorized into four distinct levels, with the most popular content strategically placed within caches situated closest to users. Furthermore, network topology and the relative positioning of nodes within the network are factored into the selection of caches for the most popular content. It is noteworthy that this method does not incorporate network coding techniques.

Proactive caching represents an innovative approach to content storage within user-proximal caches. This method leverages periods of low network traffic to pre-cache content, and it operates through distinct placement and delivery phases. Madah Ali and Niesen have proposed methodologies for implementing these phases using both centralized and decentralized approaches [4], [5]. The decentralized algorithm is presented in Algorithm 1.

During the placement phase of proactive caching, a strategic selection of content segments is stored within caches situated in close proximity to users. This pre-positioning facilitates content accessibility during peak-traffic periods, thereby maximizing opportunities for multicast transmissions. The subsequent delivery phase focuses on reducing the transmission rate by employing encoding techniques on user requests originating from various data streams. The placement phase is inherently constrained by the available cache capacity, while the delivery phase is limited by the bandwidth of the shared link. As the number of caches increases within the network, the overall transfer rate demonstrably decreases. The decentralized coded caching method, in contrast, encompasses a single placement procedure and two distinct delivery procedures. During cache placement, a random selection of M/N bits from each file is deposited within the cache. In this context, M represents the cache size relative to the total number of files (N) stored on the server. The specific delivery procedure employed is contingent upon the cache size. When the cache size exceeds one unit, the first delivery procedure is executed; otherwise, the second procedure is invoked.

The practical implementation of coded caching presents several significant challenges, including network topology considerations, asynchronous user requests, and the inherent heterogeneity of content popularity [20]. Existing network topologies for basic caching schemes were originally developed for distinct network models, such as tree networks, device-to-device communication paradigms, and hierarchical caching architectures. In contrast, primary coded caching introduces an additional layer interposed between the server and user entities. References [21], [22]. propose a two-tier hierarchical caching system. The scenarios explored by Madah Ali assume a condition where the number of files surpasses the number of users. However, references [23], [24] delve into both centralized and decentralized grouping methodologies applicable to scenarios with a larger user base. A caching system designed to store two distinct content types is presented in references [25], [26]. The first content type mirrors the design proposed by Maddah Ali, while the second type comprises linear combinations of sub-files referred to as keys. Users possessing keys retrieved from multiple caches can reconstruct the corresponding content, thereby facilitating the preservation of user privacy within the network. Notably, the first paper establishes the precondition $L \leq K/2$, whereas the second paper explores the scenario where $L > K/2$. These methods are inherently centralized and may not be well-suited for dynamic network environments characterized by frequent topological changes.

The different popularity of files is not considered in previous references and all files are considered equally popular, with a probability distribution of $p_n = 1/N$. However, it is known that some files are more popular than others so content with non-uniform distribution and multi-level popularity is introduced. The popularity of level $i$ is determined by the ratio of the number of users requesting files at level $i$ to the total number of requested files at the same level ($U_i/N_i$). The number of unpopular files is greater than the number of popular ones (i.e if $(i < j) \rightarrow N_i \leq N_j$). Various ways to partition files were introduced, the simplest method considering two levels of popularity. In the Highest-Popularity-First(HPF) caching method, the most popular files are stored in the cache with the highest popularity level. Although the expected rate is reduced by storing the most popular files for an individual cache, it is suboptimal for multiple caches.

To address this issue, the files were divided into $L$ groups based on popularity, where $N_l$ is the number of files in the l-th group, satisfying the condition $\sum_{l=1}^{L} N_l = N$. In the placement phase, files with similar popularity are grouped together, while the differences in popularity within a group are ignored. In the delivery phase, a file is randomly requested based on the probability distribution of $\{P_n\}_{n \in N}$, where $p_n$ represents the popularity of the n-th file. The delivery method $L$ times for each user group in the basic decentralized algorithm is called by the server [27]. However, this method does not guarantee optimal order. A coded transfer scheme was introduced based on the chromatic number index coding and collision graph construction when the user demand is random and follows a Zipf distribution. The expected achievable rate is obtained from the random graph [28]. An index-based placement and transfer policy with centralized coded caching is proposed by Reddy and Karamchandani. In this approach, each user is connected to $L$ caches of neighbors in a rotating manner, and users have multiple accesses to the caches. This design is optimal for fewer than four caches [29], [30].

A multi-level popularity model is proposed by Hachem et al, where users have multiple accesses to caches. The memory-sharing strategy assigns a fraction of memory to each level of popularity. For $L$ levels of popularity, $L$ subsystems are considered separately. For each level $i$ ($i \in \{1, ..., L\}$), a distinct subsystem is considered, consisting of $N_i$ files and $K$ caches. $U_i$ users are connected to each cache, and $KU_i$ users request files of level $i$. Users with different degrees of access ($d_i$) are connected to consecutive caches with symmetric periodic cycles. Each subsystem has $\alpha_i M$ memory ($\alpha_i \in [0, 1]$, $\sum_{i=1}^{L} \alpha_i = 1$) and stores non-coded chunks. The $\alpha_i$ parameters are calculated using the $M$-possible partition algorithm($H, I, J$). The algorithm is shown in algorithm 2. The description of the steps is given in references [6], [31].

This algorithm has three steps in which $\alpha_i$ values are found for all levels. Level $H$ is contained the least popular files that are not cached at all and $\alpha_h M = 0$. As a result, $KU_h$ of the requested files are answered directly from the server. $U_h$ is the number of users connected to $K$ caches that request unpopular files. $J$ is a set of the most popular files. The entire memory is allocated to it ($\alpha_j M = N_j/d_j$) and so is not needed to send data from the server. But the amount of memory for level $i$ files is reduced by them. The expression ($M - \sum_{j \in J} N_j/d_j$) in the calculation of the rate indicates this issue. The rest of the memory is allocated to files of level $i$. Finally, the total rate is calculated from the sum of the rates obtained from the three popularity levels, which is calculated in equation (1).

$$R^{MU}(M) \simeq \sum_{h \in H} KU_h + \frac{\sum_{i \in I}(\sqrt{N_i u_i})^2}{M - \sum_{j \in J} N_j/d_j} - \sum_{i \in I} d_i U_i \quad (1)$$

In this system, users connected to shared caches cannot encode with each other on the server, so a coloring-based scheme was introduced. Users with a degree of access greater

**Algorithm 2** $M$-feasible partition for all $M$ [5].

**Require:** Number of caches $K$ and parameters $\{N_i, U_i, d_i\}_i$ for $i \in 1, ..., L$.
**Ensure:** An $M$-feasible partition for all $M$.
    **for all** $i = 1, ..., L$ **do**
        $\tilde{m}_i \leftarrow (\frac{1}{k})\frac{N_i}{U_i}$
        $\tilde{M}_i \leftarrow (\frac{1}{d_i} + \frac{1}{K})\frac{N_i}{U_i}$
    **end for**
    $(x_1, ..., x_{2L}) \leftarrow sort(\tilde{m}_1, ..., \tilde{m}_L, \tilde{M}_i, ..., \tilde{M}_L)$.

    Step 1: Determine $(H, I, J)$ for each interval
        $(x_t, x_{t+1})$.
    Set $H_0 \leftarrow \{1, ..., L\}, I_0 \leftarrow \emptyset, J_0 \leftarrow \emptyset$.
    **for** $t \leftarrow 1, ..., 2L$ **do**
        **if** $x_t = \tilde{m}_i$ for some $i$ **then**
            promote level $i$ from $H$ to $I$
            $H_t \leftarrow H_{t-1} \setminus \{i\}$
            $I_t \leftarrow I_{t-1} \cup \{i\}$
            $J_t \leftarrow J_{t-1}$
        **else if** $x_t = \tilde{M}_i$ for some $i$ **then**
            Promote level $i$ from $I$ to $J$
            $H_t \leftarrow H_{t-1}$
            $I_t \leftarrow I_{t-1} \setminus \{i\}$
            $J_t \leftarrow J_{t-1} \cup \{i\}$
        **end if**
    **end for**

    Step 2: Compute the limits of the intervals as $[Y_t, Y_{t+1})$.
    **for all** $t \in \{1, ..., 2L\}$ **do**
        $Y_t \leftarrow x_t \cdot S_{I_t} + T_{J_t} - V_{I_t}$
    **end for**
    $Y_{2L+1} \leftarrow \infty$

    Step 3: Determine the $M$-feasible partition for all $M$.
    **for all** $t \in \{1, ..., 2L\}$ **do**
        Set $(H_t, I_t, J_t)$ as the $M$-feasible partition of
        $M \in [Y_t, Y_{t+1})$
    **end for**

than one are grouped together to avoid encoding between requests of the same cache. To implement this design in practice, coding between two different groups with different caches is possible. In this paper, queues are created based on the number of network caches for users with access to two caches. With the addition of each cache to the network, a new queue is added.

## III. INTEGRATING CCN WITH DCC WITH MULTI-LEVEL POPULARITY AND ACCESS

This work investigates a Content-Centric Network (CCN) scenario involving a server and a collection of hosts. All hosts possess content libraries containing identical files of fixed size (F bits). The system model leverages decentralized coded caching, which factors in content with varying popularity levels and user access disparities. This approach is grounded in the findings presented in [6], [31], which demonstrate that partitioning content based on the number of lowpopularity levels yields optimal results. In this context, content categorized into two popularity levels is strategically cached at $K$ caches situated in close proximity to end users ($L = 2$). The following assumptions are established initially:

- The files are assumed to have equal sizes and are divided into equal chunks.
- The number of caches is a multiple of the highest degree of access. The capacity of the caches is considered equal and is not related to each other. Uncoded data is stored in the content store and caches close to the user.
- The system is designed for multiple users, with varying degrees of access to the cache taken into account. User requests are processed concurrently.

Users who request files with a popularity level of $i$ have access to $d_i$ caches. Caches and files are divided into two groups based on the highest degree of access ($d_{max} = 2$). Popular and unpopular files are requested by users with access levels one and two, respectively. The basic decentralized coded caching is used to store file chunks in caches. In the content placement phase in caches, content chunks are cached randomly and independently, using the $M$-feasible partition algorithm. A certain amount of memory is assigned to each level of popularity with this method, and for each level, a fraction of memory is determined based on the capacity of caches, the amount of user access to caches, and the number of levels of popularity. The system is multi-user, and user requests are considered simultaneously. Chunks are selected based on cache capacity for each popularity level and stored without coding. Each level of popularity is considered as a separate subsystem with less memory, and content delivery is done for each level separately. To integrate this method into the content-centric network, FIFO queues are used separately for each popularity level. In the basic decentralized coded caching, caches are single-user and use a single queue. During queuing, interests are searched from other caches with different chunks. Encoding is only possible between users connected to different caches. When multiple users are connected to the same cache, coding between their requests is not possible.

This paper proposes a novel multi-level access scheme for caches within a coded caching framework. The scheme caters to scenarios where users possess access to a single cache. Under such conditions, the number of queues maintained by the server is equivalent to the maximum access degree, with a dedicated queue designated for each color. Notably, the transmission rate across the network is reduced by preferentially transmitting coded packets constructed from data retrieved from two queues with distinct colors. In traditional decentralized coded caching, each coding operation necessitates a comprehensive search of the entire queue, leading to potential processing inefficiencies. To address this limitation, the proposed scheme introduces a color-based separation strategy during the initial packet placement phase. Packets are categorized and deposited into separate queues based on their assigned color. This approach eliminates the need for exhaustive queue searches during the encoding process, as packets are retrieved from the beginning of their respective color-designated queues.

### A. The proposed method for the maximum degree of access

- Groups (or colors) are created based on the maximum degree of user access to caches ($d_{max} = 2$).
- First-In-First-Out (FIFO) queues are created for each group on the server. Their number(Num) is determined by the equation (2).

$$\text{Num} = K/maxd_i. \quad (2)$$

$K$ represents the number of caches, and $maxd_i$ represents the maximum degree of popularity access at level $i$.
- Interests sent from a cache unit are placed in a queue.
- Interests are coded in a queue of the same color. This has the following benefits:
  - Coded data packets are easily decoded in sinks.
  - To avoid interference between user requests connected to a cache, queues with different colors are created, and encoding is applied between two chunks of the same color.

### B. An example of the proposed method

A network with $d_{max} = 2$ is shown in Fig 1. The files consist of 12 chunks with two levels of popularity and are divided into two parts based on $d_{max}$. The first half of the chunks of each file are randomly and uniformly stored in the orange caches (e.g.,{1, 2, 4, 5}), while the green caches are filled with the second half of the file chunks (e.g., {8, 9, 10, 11}). Two types of users request content, which is classified according to their level of popularity. Users with access levels one and two request popular and unwanted files, respectively.

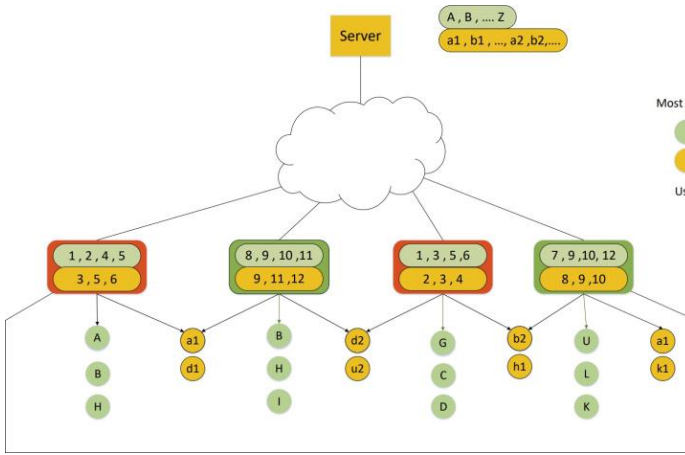

Fig. 1: Coded caching for Multi-Level Popularity And Access with $K = 4$ caches and $L = 2$ levels is illustrated. The caches are colored by two colors and files are partitioned and colored by the same colors. Users ($U_i$, $i \in \{1, 2\}$)with two access levels is connected to caches.

### C. Coding Gain

- **Users with access level $d_i = 1$** : The server's queue count is determined by the number of groups (colors). Users connected to the same cache are placed in a queue. Coding between two caches is shown in Fig 2(a). With this approach, it takes only 12 transmissions to delivery two files, whereas CCN requires 24. Coding gain is calculated as the ratio of uncoded to coded transmissions from reference [32], which gives a double advantage here.
- **Users with access level $d_i = 2$** : The server has two types of queues. Users connected to different caches are coded together with the same color. Coding between two caches is shown in Fig 2(b). With this method, sending two files requires only 8 transmissions, resulting in a coding gain of 3.

## IV. INTEGRATE DECENTRALIZED CODED CACHING WITH CCN

Incorporating decentralized coded caching into CCN necessitates numerous changes to the architecture, which are outlined below.

### A. Changes in interest

In Content-Centric Networks, interests are sent to the server from various receivers and are identified by a one-bit field with a value of 0, while data packets are identified with a value of 1. However, in decentralized coded caching, interests from multiple receiver nodes are combined or coded together on the server. To implement this in CCN, changes are required in the interest and data packets. The files are divided into different levels, each containing $m$ chunks. $C_{1..n} = C_1, C_2, ..., C_n$ are the names of the files involved in the XOR operation and stored in the header of the data packet. The maximum number of coded chunks is equal to $n$. More details about these changes are provided below.

- **Interest** ($C_i$, P, Color, $S_{Cache}$, $S_{New}$): When interests are generated, the value of $C_{1..n} = C_i$ represents the name of the packet requested by the user from the i-th cache. P indicates the popularity of the content name. The files stored in the cache are determined by color (for example, orange color is included chunks of the first half of the files, and green color is included second half of the file). The set of cached chunks is denoted by $S_{Cache} : S_1, S_2, ..., S_n$, which at the time of sending the interest from sink $i$, is equal to $S_i$. The index of the coded chunks is placed in the $S_{New}$ vector, which initially has an empty value in the receiver and is filled in the sending path or the server after the encoding operation.
- **Data** ($C_{1..n}$, P, Color, $S_{Cache}$, $S_{New}$): In the server, the coded chunks are assembled in a data packet whose number, for users with an access degree of one, ranges from a minimum of one to a maximum of the number of cache colors, and for users with the maximum access degree, from one to $k/max_{d_i}$. The CCN packet has a field called HopLimit, which is similar to the TTL field in an IP-based network. This field is used to transfer the packets from the server. When there are no other requests from other sinks for encoding in several time slots, the packet is returned to the receiver before the timeout.

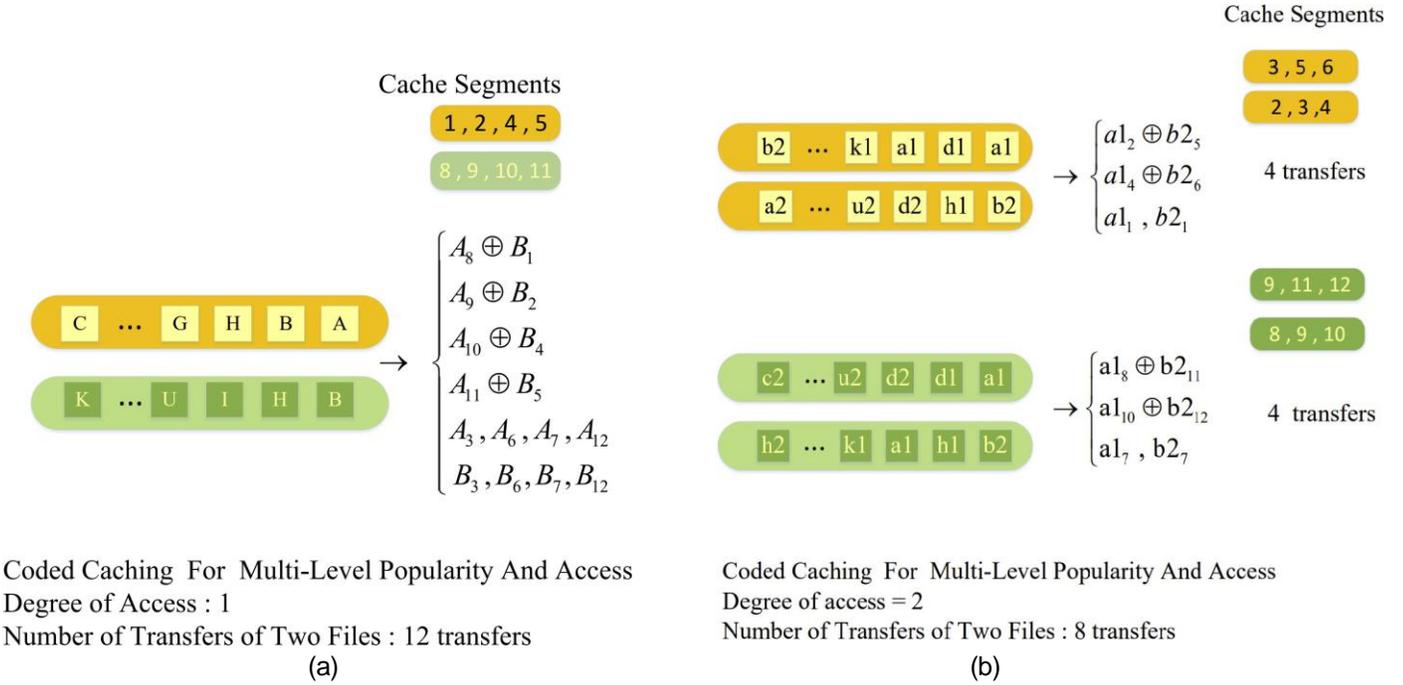

Fig. 2: Users with different degrees of access. (a) Users with $d_i = 1$ (b) Users with $d_i = 2$.

*B. Interest processing*

When an intermediate node interface receives an interest, the CS is searched for the content name. If the CS has all $m$ chunks, data messages are generated and sent downstream. If the CS has $x$ chunks of content ($|x| < m$), a new interest is created. For example, a cache that contains the first half of a file (e.g., {1, 2, 4, 5}) requests content A(popularity A = 1). Along the path, it finds chunks {3, 7, 8} in the CS. These chunks are stored in the $S_{New}$ field and new interest is sent to the server as $(A, P : 1, \{, 2, 4, 5\}, \{3, 7, 8\})$. This is illustrated in Fig 3. At the server, the queue is selected based on the content popularity and $S_{Cache}$. If $S_{New}$ is not empty, then $S_{Cache} = S_{Cache} \cup S_{New}$. Based on the new $S_{Cache}$, the basic decentralized coded caching procedure is executed.

*C. Data packet processing*

When receiving coded data, it is searched for content names $C_{1..n} = C_1, C_2, ..., C_n$ in the PIT. To minimize search time, each record in the PIT is compared with all names in the coded data and a new chunk is added to the PIT with each search. Deleting a record from the PIT is dependent on user access and is equal to the ratio of the number of chunks to the user's access level. Once the set of requested chunks is complete, it will be removed from the PIT.

*D. Changes in CS*

The uncoded data is stored in the content store, and popular data has a higher priority for storage. Additionally, the LRU (Least Recently Used) method is implemented as a content storage replacement policy. If the content store is full, the oldest unused record is selected for replacement by the LRU method.

*E. Recoding in intermediate nodes*

After encoding, some data is sent back to the sink node in uncoded form. To increase throughput, uncoded data from different sink nodes is encoded in intermediate nodes. For content with popularity=1 and $d_i = 1$, each data is encoded using the primitive approach with packets from other sink nodes. Unpopular content with $d_i = 2$ is encoded only with data from the same half-file (same color), thus avoiding encoding data from the shared cache.

V. EVALUATION

The simulation is implemented using MATLAB software. A Content-Centric Network with 20 nodes is considered. 150 files are stored on the server and each file is divided into 12 chunks of equal size. Four caches of the same size are considered close to the user. The caches have no connections with each other and their capacity is 360 MB. The links in the network have different capacities. The capacity of the content store is estimated at 150 packets(150 MB). The simulation parameters are shown in Table I.

Files are grouped by the highest degree of user access, with a maximum degree of access of 2 ($d_{max} = 2$). An unlimited number of users with different degrees of access ($d_i \in \{1, 2\}$) are connected to the cache. The number of user requests is determined randomly using a Poisson distribution function, with $\lambda = Landa_i$ for each level of popularity ($i$), and the number of users for each sink node is determined using the equation (3).

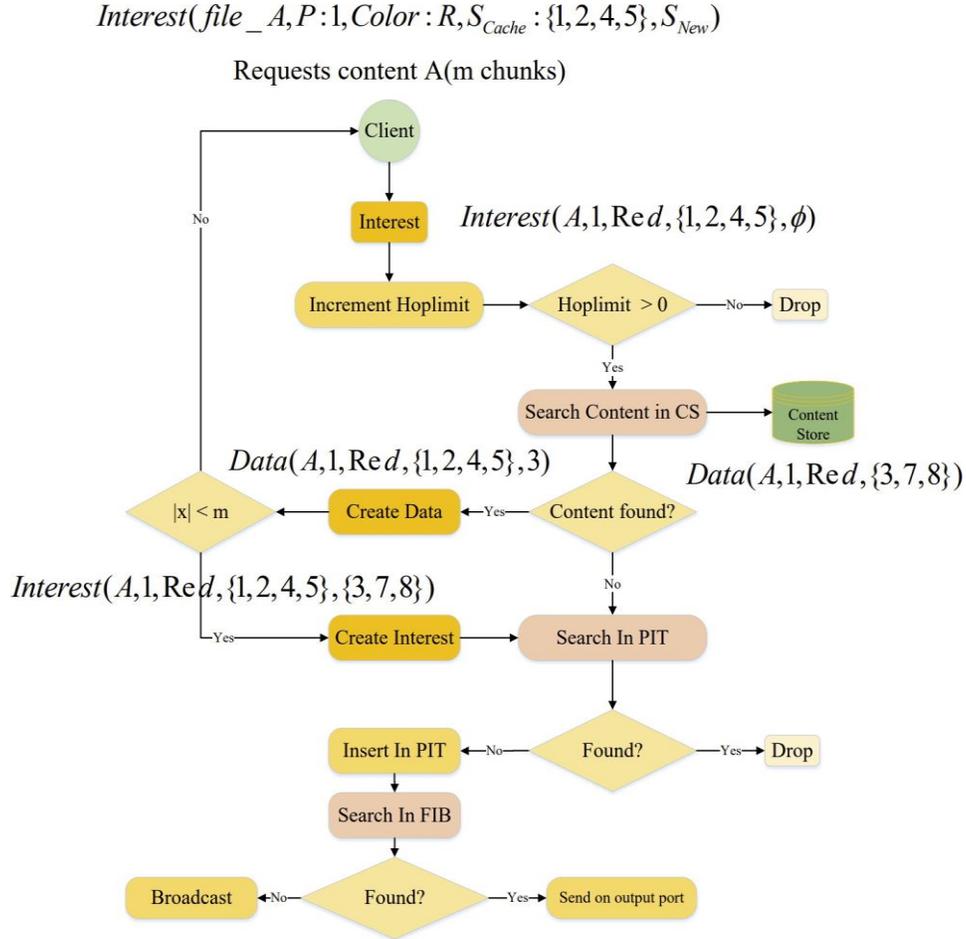

Fig. 3: Interest packet processing

TABLE I: Problem parameters

| Parameter | Value | Description |
|---|---|---|
| Capacity of the final caches | 30 Files | 360 MegaBits |
| Capacity of content store(CS) | 150 Packets | 150 MegaBits |
| Number of files | 150 | |
| file size | 12 MegaBits | Same size |
| Number of file chunks | 12 | Same size |
| Size of the chunks | 1 MegaBit | |
| Link capacity | 50 MegaBits | 50 interests |
| Size of the time slice | 1 second | |
| Maximum access level | 2 | |
| Simulation time | 60 | time slices |

Number Of User$_i$ = $poissrnd(\lambda, [1, \text{Number Of Sinks}])$ (3)

The improvement of the proposed method in terms of throughput, delay, and cache hit rate is observable in all the diagrams compared to the conventional CCN method. In the coded caching method, the server XORs the interests, resulting in a reduction in the number of transmissions in the network. In each time slice, more packets can pass through the network link, which increases the network's throughput. Additionally, the rate of throughput improvement is accelerated by re-encoding the uncoded packets along the route.

The comparison of the proposed method's throughput with content-centric networks and the NetCodCCN protocol from reference [11], using MATLAB software, is shown in Fig 4. The value of $\lambda$ (the number of user requests) varies from 4 to 10, and with the increase in $\lambda$, the throughput increases, nearly doubling at $\lambda = 8$. When comparing this with the RLNC-based method, the improvement in throughput is clearly noticeable. The comparison of the three methods in terms of average throughput is shown in Table II.

Various cache placement policies such as LRU, LFU, and Random have been integrated with the proposed method and CCN in Fig 5(a). By XORing packets in the network for each flow, fewer packets need to be sent through the channel, thereby allowing the network to respond to more users based on the limited link capacity. This increases the network's throughput as the number of users increases. In the CCN method, as the number of users increases, the number of requests answered is limited by the link capacity, and therefore, throughput cannot increase as it does with the coded method. A comparison of the two methods at $\lambda$ values ranging

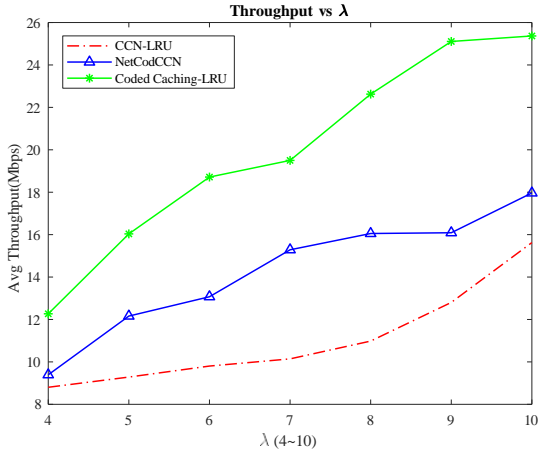

Fig. 4: Comparison of the throughput of the proposed method with the Content-Centric Networks and the NetCodCCN protocol

TABLE II: Average throughput

| $\lambda$ | CCN | NetCodCCN | $\lambda$ (Min,Max) | Coded Caching |
|---|---|---|---|---|
| 4 | 8.804 | 9.393 | 1 , 3 | 12.27 |
| 5 | 9.286 | 12.16 | 2 , 3 | 16.04 |
| 6 | 9.804 | 13.07 | 2 , 4 | 18.71 |
| 7 | 10.14 | 15.29 | 3 , 4 | 19.50 |
| 8 | 10.98 | 16.05 | 3 , 5 | 22.63 |
| 9 | 12.80 | 16.09 | 3 , 6 | 23.39 |
| 10 | 15.63 | 17.96 | 4 , 6 | 24.32 |

from 3 to 12 shows nearly a threefold increase at $\lambda = 9$.

The comparison of the two methods in 5(b) is based on different content store (CS) sizes of 50, 100, 150, 200, and 250 with a fixed number of users ($\lambda = 7$, generated by a Poisson distribution). As the cache size increases, fewer coded data packets are sent from the server to the requesting node, or in other words, uncoded data packets are sent from intermediate nodes to the receiver. This highlights the impact of coding on increasing throughput. The improvement at a cache size of 100 shows nearly a threefold increase.

End-to-end delay refers to the average time between the generation of a packet at the source node and its successful delivery at the destination. Delay in CCN with the proposed method is compared at $\lambda$ values ranging from 3 to 11 in Fig 5(c). As the number of users in the network increases, the traffic in the network increases, and the server queue becomes busier. This leads to increased response delays in the network. However, overall, since the number of coded packets transmitted is fewer than the uncoded ones in the conventional CCN, more requests can be answered, ultimately reducing delay. At $\lambda = 5$, the average delay is halved. Fig 5(d) shows the lower delay (Almost half the amount) of coded data compared to uncoded data. The number of users is kept constant. As the content store size increases, data is sent from the middle of the network to the receiver, reducing the overall delay.

Popular content is prioritized for storage in intermediate caches, leading to an increased cache hit ratio as user demand increases. The average cache size is set to 150 MB, and the $\lambda$ parameter varies from 3 to 12. Fig 5(e) shows that the cache hit ratio triples at $\lambda = 8$.

Fig 5(f) provides a comparative analysis of cache hit rates at different content store (CS) capacities. In this experiment, the $\lambda$ parameter is set to 7. The CS capacity varies in four scenarios, including configurations of 50, 100, 150, 200, and 250 MB. As the CS capacity increases, a larger volume of data can be stored in the network. Consequently, subsequent user requests are more likely to be fulfilled by intermediate nodes, resulting in an increased cache hit rate. Notably, the cache hit rate nearly triples when the CS capacity is set to 100 MB.

## VI. CONCLUSION

Caching popular content represents a well-established technique for enhancing network throughput and minimizing user-experienced delays. This approach entails replicating frequently accessed content within a distributed cache network, strategically positioned for ease of user access. Consequently, requested content can be delivered to users with demonstrably reduced latency. Coded caching offers the additional benefit of fulfilling multiple user requests with a single coded data transmission. However, to achieve this efficiency, effective management of both caching and coding functionalities is paramount. A key challenge lies in the infeasibility of en-coding user requests associated with a singular cache. Additionally, searching for each individual user request within the queue can incur significant processing costs. This paper proposes a novel solution specifically designed to achieve optimal coding efficiency while minimizing associated costs. The conducted evaluation demonstrably reveals that this approach leads to substantial improvements in network throughput, user-experienced delay, and overall cache hit ratio when compared to conventional Content-Centric Network(CCN) architectures.

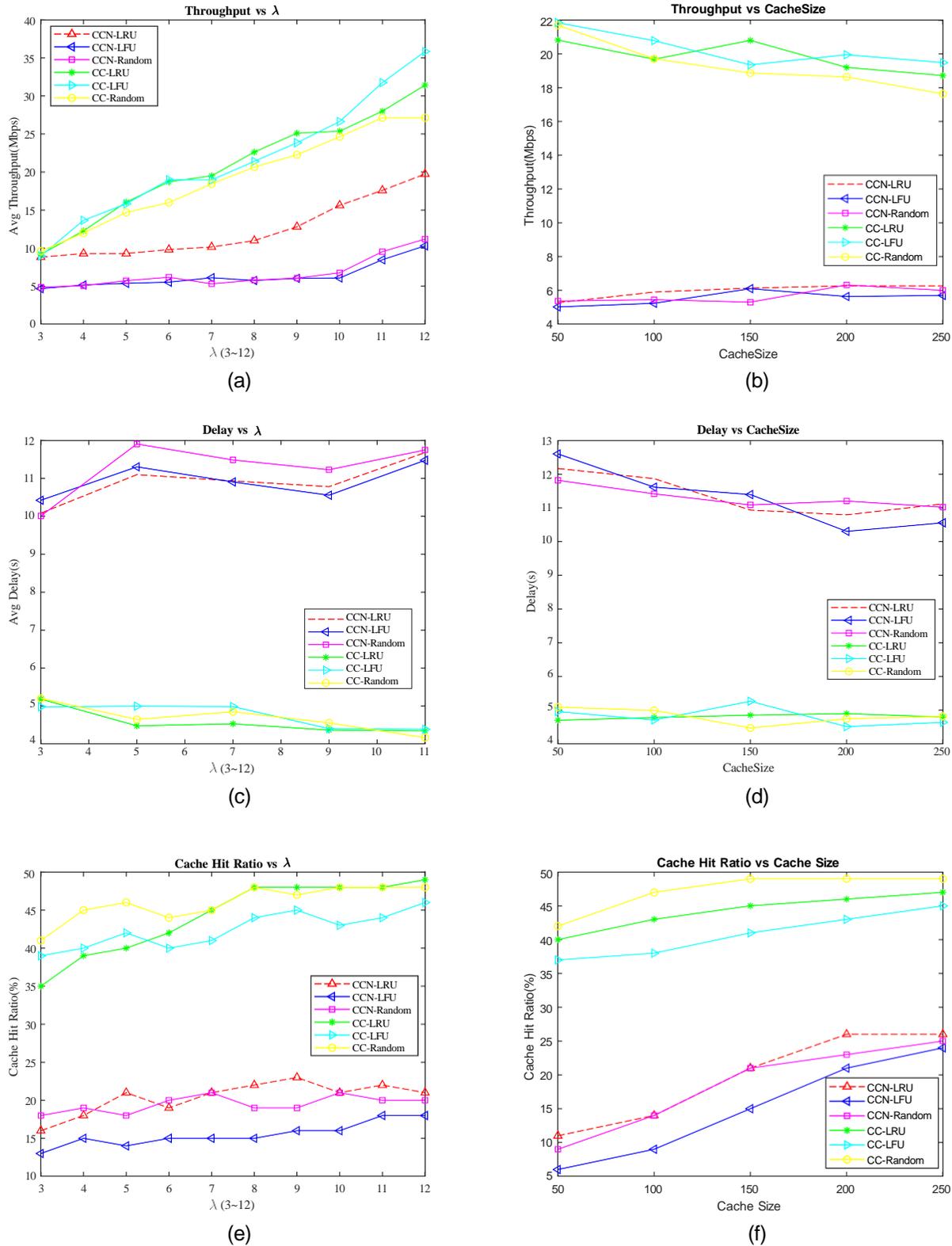

Fig. 5: Comparing the proposed method with CCN with different cache-replacement policies. (a) Average throughput vs $\lambda$. (b) Average throughput vs different cache size. (c) Delay vs $\lambda$. (d) Delay vs different cache size (e) CacheHit vs $\lambda$. (f) Cache Hit vs different cache size.